# Attention-Guided Multi-Scale Local Reconstruction for Point Clouds via Masked Autoencoder Self-Supervised Learning

Xin Cao[a], Haoyu Wang[a], Yuzhu Mao[a], Xinda Liu[a], Linzhi Su[a,] and Kang Li[a]

[a]*School of Information Science and Technology, Northwest University, Xi'an, Shaanxi, China*

**Abstract**

Self-supervised learning has emerged as a prominent research direction in point cloud processing. While existing models predominantly concentrate on reconstruction tasks at higher encoder layers, they often neglect the effective utilization of low-level local features, which are typically employed solely for activation computations rather than directly contributing to reconstruction tasks. To overcome this limitation, we introduce PointAMaLR, a novel self-supervised learning framework that enhances feature representation and processing accuracy through attention-guided multi-scale local reconstruction. PointAMaLR implements hierarchical reconstruction across multiple local regions, with lower layers focusing on fine-scale feature restoration while upper layers address coarse-scale feature reconstruction, thereby enabling complex inter-patch interactions. Furthermore, to augment feature representation capabilities, we incorporate a Local Attention (LA) module in the embedding layer to enhance semantic feature understanding. Comprehensive experiments on benchmark datasets ModelNet and ShapeNet demonstrate PointAMaLR's superior accuracy and quality in both classification and reconstruction tasks. Moreover, when evaluated on the real-world dataset ScanObjectNN and the 3D large scene segmentation dataset S3DIS, our model achieves highly competitive performance metrics. These results not only validate PointAMaLR's effectiveness in multi-scale semantic understanding but also underscore its practical applicability in real-world scenarios.

## 1. Introduction

In recent years, the rapid advancement of deep learning technologies has sparked revolutionary changes across various fields, particularly in 3D technologies. As a core data format, point cloud provides an accurate digital representation of objects in three-dimensional space, serving a variety of practical 3D applications such as autonomous driving, virtual worlds and robotics. Since the introduction of PointNet[1], point cloud data processing has entered a new era, laying the foundation for the application of deep learning in this domain.

However, Point cloud data analysis is primarily dependent on supervised learning[2]. This approach necessitates large-scale labeled datasets for learning the mapping between input data and labels. Due to the unordered and irregular nature of point cloud data, manual labeling is challenging and error-prone. These limitations impact the model's generalization ability and scalability. To mitigate these issues, self-supervised learning has been introduced, aiming to decrease reliance on labeled data and to improve the model's generalization capabilities.

Self-supervised learning, a powerful unsupervised learning paradigm[3], enables models to uncover and utilize intrinsic structures and features within datasets without labeled data. This field is primarily achieved through two methods: generative learning and contrastive learning. Generative learning, which focuses on simulating data distributions, utilizes models such as Generative Adversarial Networks (GANs)[4] and autoencoders[5] to reconstruct input data or predict its potential future states. Contrastive learning (CL)[6], on the other hand, employs discriminative methods, using similarity metrics to bring closely related samples closer together while pushing dissimilar samples apart. These strategies have been significantly successful in computer vision and natural language processing.

Nevertheless, progress in self-supervised learning in the three-dimensional point cloud domain has been relatively slow. The unordered nature and spatial irregularity of point cloud data present challenges for developing robust feature learning methods. The unique properties of point clouds require innovative approaches to effectively capture and utilize their spatial information. Inspired by significant advancements in natural language processing (NLP) and image analysis, many researchers have begun to shift their focus toward self-supervised learning for 3D point clouds. This shift has facilitated the development of methods specifically designed for self-supervised learning in point clouds, including contrastive learning-based approaches[7] and generative learning methods[8]. In the current research landscape, techniques based on Masked Autoencoders have been widely applied to the processing of three-dimensional point clouds[9]. These methods follow a core mechanism that involves performing random masking operations on the input point cloud. Only the unmasked subset of the point cloud is fed into the encoder. The model's objective is to reconstruct the masked portions of the point cloud through a decoding process. In this way, the model is forced to infer the complete three-dimensional structure from limited observations. This process helps the model learn deep feature representations of point cloud data. Current model architectures often focus on performing reconstruction tasks at the higher layers of the encoder, failing to fully utilize the information from the lower layers. This approach usually results in the model being unable to effectively extract rich local details and geometric structure information from low-level features when processing input data. In these models, low-level local feature interactions are primarily used for the computation of activation functions, rather than directly participating in the reconstruction tasks. Consequently, the crucial role of these local features in capturing fine structures is not fully realized. This strategy overlooks the importance of low-level features in capturing local details and geometric structures, which are essential for executing accurate point cloud processing tasks.

The multi-scale local reconstruction technology has garnered significant attention in the field of image processing for its potential to enhance feature expression and processing accuracy. Recent research indicates that multi-scale analysis effectively captures both local details and global context information of images. Wang et al. [10] proposed a self-supervised learning method, LocalMIM, which enhances semantic understanding of features through multi-scale reconstruction at various local layers of the encoder. This research highlights the potential of multi-scale reconstruction in improving image representation. Concurrently, Chen et al. [11] introduced CLIT, a novel arbitrary scale super-resolution method that integrates attention mechanisms and frequency coding technology, significantly improving image super-resolution performance. Drawing inspiration from these studies, we aim to incorporate the concept of multi-scale local reconstruction into point cloud data to bolster the representational capabilities of point clouds. We propose PointAMaLR, a self-supervised learning method that enhances feature expression and processing accuracy through attention-guided multi-scale local reconstruction. The PointAMaLR model accounts for the complex interactions between patches necessary for reconstruction tasks to infer target signals and is applied across multiple local layers, both lower and higher, facilitating multi-scale semantic understanding of input data. Additionally, to further refine feature representation, a Local Attention (LA) module is introduced at each local layer to enhance the semantic understanding of features.

Extensive experiments on public datasets show that our approach achieves competitive results and is currently the most advanced on two variants of ScanObjectNN, a challenging real-world dataset. We summarize the main contributions of this paper as follows:

1. We propose PointAMaLR, an innovative self-supervised learning framework designed to achieve hierarchical representation learning of point clouds through local attention mechanism-guided multi-scale local reconstruction tasks.
2. We designed a Local Attention (LA) module to guide the reconstruction of features at both low and high levels, enhancing the model's understanding of local details and global structures.
3. The experimental results demonstrate that PointAMaLR achieves outstanding classification accuracy on standard datasets such as ModelNet and ShapeNet[12], and also delivers high-quality results in point cloud reconstruction tasks. Moreover, PointAMaLR exhibits competitive high performance on the ScanObjectNN real-world dataset, further validating its potential for generalization across different environments.

## 2. Author Artwork

*2.1. Masked autoencoders*

In the fields of natural language processing and computer vision, Masked Autoencoders (MAE)[13]have demonstrated their powerful self-supervised learning capabilities by introducing masked elements into the data, effectively simulating the mechanisms of self-supervised learning (SSL). For example, the BERT[14] model in natural language processing utilizes masked language modeling to predict the masked words in a text. In the computer vision domain, similar approaches such as MAE and SimMIM[15] mask parts of an image, forcing the model to infer and reconstruct the obscured sections from the visible content. These successful cases have sparked research into self-supervised learning for three-dimensional point cloud data processing. Point-BERT[16], as a pioneer in this field, was the first to use MAE for pre-training small point cloud datasets. Following this, Point-MAE[17] employed chamfer distance to reconstruct the masked portions of point clouds, while MaskPoint[18] designed a specific decoder to identify and process small masked point sets. In recent research, PointGame[19] has developed a geometrically adaptive mask autoencoder for extracting key features from point clouds. At the same time, PCP-MAE[20] introduces a self-supervised learning framework that focuses on predicting the center of the masked point cloud, thus enhancing the feature extraction process. These methods share a common strategy: randomly masking certain parts of the point cloud data and relying on the autoencoder's ability to reconstruct these obscured sections, thereby enhancing the model's learning and feature extraction capabilities. Through this approach, self-supervised learning for point cloud data has significantly advanced, opening new possibilities for three-dimensional data processing.

*2.2. Point Cloud Representation Learning*

3D point clouds are different from 2D images, which are arranged neatly on a regular grid, while the former are distributed irregularly in three-dimensional space. This irregularity makes it challenging for traditional convolutional neural networks (CNNS) to reliably extract features from these point clouds. To address this challenge, researchers have proposed a variety of solutions. Projection-based method [21] converts the disordered point cloud into 2D image by projecting the point cloud onto the image plane, and then uses 2D CNNs to extract features, and finally fuses these features to form the final output features. The voxel-based approach [22] solves the disorder problem by converting the point cloud into 3D voxels and applying 3D convolution to extract features, but this can lead to a significant increase in computing and memory requirements with increasing resolution, as the number of voxels grows exponentially. The point-based approach [1, 23] deals directly with the convergence of points embedded in a continuous space.

PointNet[1] is a pioneering network that operates directly on unordered points and passes information through pooling operators. PointNet++[23] Based on PointNet[1], a hierarchical feature learning method for recursively capturing local geometric structures is proposed. DGCNN[24] connects point clouds into local graphs and dynamically computes these graphs to extract geometric features.

In natural language processing (NLP) and 2D computer vision (CV) tasks, transformer structures have flourished for a long time. Inspired by the successful application of converters in NLP [25] and image regions[13], many 3D vision models have emerged. In the field of 3D computer vision, converter-based approaches typically employ an encoder-decoder framework with multiple layers of self-attention and point-by-point, fully connected layers. The self-attention mechanism is particularly well suited for dealing with 3D point clouds because it is able to naturally process a set of points with location information. Researchers such as Zhao et al. [26] have developed point transformation layers that utilize self-attention for various 3D scene understanding tasks, including semantic segmentation, object partial segmentation, and object classification. SAPFormer[27] further enhances the feature propagation process by introducing a shape-aware attention mechanism that strengthens both local and global context modeling, yielding improved performance across several point cloud benchmarks. Similarly, LCASAFormer[28] introduces a cross-attention mechanism to enhance the integration of local and global features, achieving superior performance in 3D object detection and segmentation tasks. Transformer-based methods offer notable advantages, such as high parallelism, the ability to capture long-range dependencies, and minimal inductive bias. Their global modeling capability is particularly beneficial for understanding the holistic structure of point cloud data.

*2.3. Self-Supervised Learning*

In the representation learning of three-dimensional point clouds, self-supervised learning (SSL) has emerged as a key research area, with various studies demonstrating its effectiveness. The main appeal of SSL lies in its ability to learn by mining the intrinsic structures and features of the data without the need for external annotations or supervision. Through carefully designed pre-training tasks, SSL encourages models to enhance themselves, typically involving the reconstruction of input point clouds after undergoing some transformation based on the encoded latent vectors, such as through rotation, deformation, partial rearrangement, or masking. For example, CrossPoint[29] serves as a cross-modal contrastive learning method, obtaining rich self-supervised learning signals by comparing point clouds with their corresponding 2D rendered images. Huang et al. introduced the STRL[30] method, which fosters a paradigm of unsupervised learning through strategic interaction between an online network and a target network to enhance learning outcomes. Significant progress has also been made in the research field of autoencoders. Wang et al. proposed OcCo[31], a novel encoder-decoder structure specifically designed to effectively reconstruct partially occluded point cloud data, thus enhancing point cloud processing capabilities when dealing with visibility challenges. Point-BERT is a method that uses a Masked Point Modeling (MPM) task to pre-train point cloud transformers, leveraging discrete variational autoencoders (dVAE) to generate discrete point labels rich in local information. Nonetheless, the reliance of Point-BERT on dVAE pre-training introduces additional complexity. To simplify this process, Point-MAE focuses on masking tasks to streamline pre-training and improve model performance through subsequent downstream tasks. This approach enables the model to learn point cloud representations more efficiently by simplifying the pre-training steps, providing a strong foundation for subsequent tasks.

## 3. Method

The overall architecture of PointAMaLR is shown in Figure 1, detailing our multi-scale local reconstruction framework and the integrated LA attention mechanism in the model design.

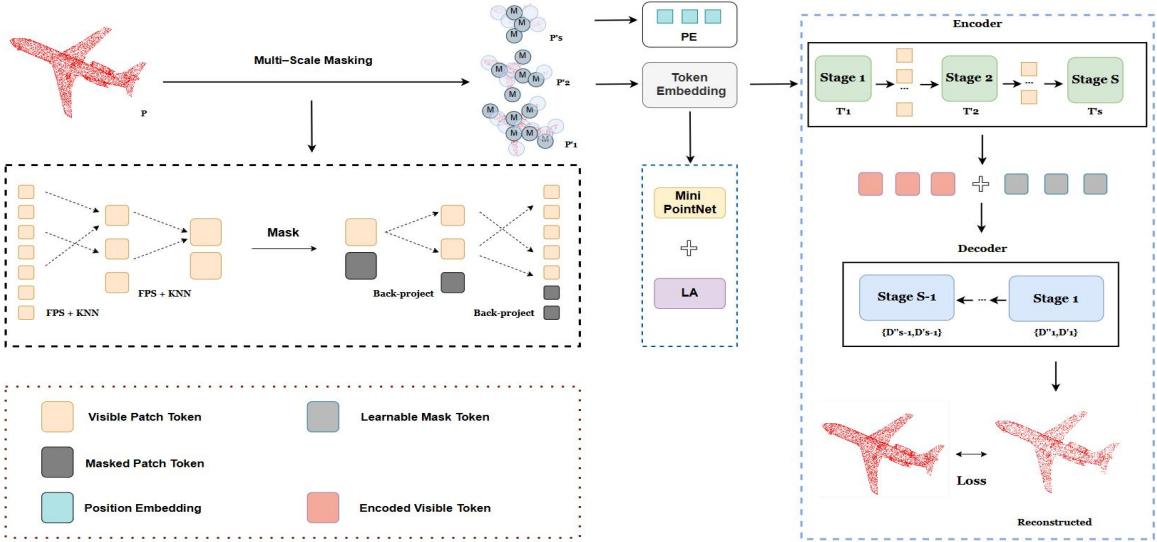

Fig. 1. Framework of PointAMaLR. Firstly, we apply a multi-scale masking strategy to the original point cloud, and then, at the highest level after the multi-scale mask, the visible token is sent to the multi-stage encoder through an embedded processing part with a local attention module (LA) to capture the multi-scale semantics of the point cloud. Finally, we reconstruct the masked part of the input point cloud through a multi-stage decoder module that integrates the self-attention layer and the prediction head.

### 3.1. Masking Strategy

To construct a mask autoencoder for multi-scale learning, we learned $S$ - scale techniques to encode point clouds[32]. Each scale within the $S$ - scale framework corresponds to a distinct number of points, necessitating the adaptation of the standard encoder to an S - class architecture. Consistent with prior research, after embedding point clouds into discrete point markers, we apply random masking for the purpose of reconstruction. Recognizing that the block portions of 3D shapes are more likely to preserve a more complete fine-grained geometry, and that unmasked locations should ideally be shared across all scales to enable the encoder to perform coherent feature learning, it is imperative for the unmasked regions in a multi-scale architecture to exhibit consistency across different scales, not just within a single scale. This is particularly pertinent for points that are irregularly distributed. Consequently, we initially develop an $S$ - scale representation of the input point cloud and then back-project the final $S$ - scale random mask to an earlier scale. This method reduces the detrimental effects of excessive fragmentation in the visible portion, which can be induced by fully random masks, thereby enhancing learning performance by minimizing the negative impact of such fragmentation.

### 3.2. Different scale representation

We represent the input point cloud as $P$ and consider it as the 0 th-level scale. For each first-level scale $i$ where $0 \leq i \leq S$, we employ the farthest point sampling (FPS) method to downsample from the $(i -$

1) $-th$ scale, obtaining the target point $P_i \in \mathbb{R}^{N \times 3}$ for scale $i$. Following previous research methods, we use $k$ -nearest neighbors (KNN) to aggregate k neighboring points for each target point, simultaneously obtaining the neighbor index $I_i \in \mathbb{R}^{N_i \times k}$. Through continuous downsampling and grouping, we derive the $S$ - scale representation of the input point cloud $\{P_i, I_i\}$ for $S = 1$, where the number of points $N_i$ decreases progressively, and the hierarchical relationships between scales are recorded in $I_i$. The specific formulas are as follows:

$$P_i = FPS(P_{i-1}), P_i \in \mathbb{R}^{(N_i \times 3)} \quad (1)$$
$$I_i = KNN(P_i, P), I_i \in \mathbb{R}^{(N_i \times k)} \quad (2)$$

*3.3. Back-projecting*

For the points at the final $S$ - scale, denoted as $P_S$, we apply a random masking procedure with a maximum mask ratio of 60%. The remaining visible points are denoted as $P'_S$. Given that parts of 3D shapes tend to retain more complete fine-grained geometry, it is beneficial for unmasked positions to be consistent across all scales to facilitate coherent feature learning within the encoder. Therefore, we back-project from the final S - scale to the lower scales.

For any $1 \le i \le S$, we retrieve all $k$ nearest neighbors of $P'_{i+1}$ using the index $I_{i+1}$, assigning them as the visible points $P'_i$ for that scale, while masking the remaining positions. Through this recursive back-projection, we obtain visible and masked positions for all $S$ - scale, represented as $\{P'_i, P''_i\}_{i=1}^S$, where $P'_i \in \mathbb{R}^{N'_i \times 3}$, $P''_i \in \mathbb{R}^{N''_i \times 3}$ and $N_i = N'_i + N''_i$.

*3.4. Token Embedding*

Indexed by $I_1$, we utilize a lightweight PointNet to extract and fuse the features of every seed point from $P'_1 \in \mathbb{R}^{N_1 \times 3}$ with its k nearest neighbors, it converts each block of the point cloud into a series of embeddings. Then, we derive the initial point tokens $T'_1 \in \mathbb{R}^{N_1 \times C_1}$ for the encoder's first stage, which embeds $N_1^e$ local patterns of the 3D shape. In the transition from the $(i-1)-th$ stage to the $i-th$ stage (where $1 \le i \le S$), we combine $T'_{i-1} \in \mathbb{R}^{N_{i-1} \times C_{i-1}}$ to produce the downsampled point tokens for the $i-th$ stage. This process employs multi-layer perceptron (MLP) layers along with max pooling to aggregate the $k$ tokens nearest to $P'_i$ as indexed by $I_i$, resulting in the output $T'_i \in \mathbb{R}^{N_i \times C_i}$. Thanks to our multi-scale masking approach, the combined $T'_i$ aligns with the corresponding visible parts of $T'_{i-1}$, facilitating consistent feature encoding across varying scales.

In order to achieve our multi-scale local reconstruction goal, the designed LA module is embedded in embedding process. We embed the module after the first conv and after the second conv, as shown in Figure 2. The reason for embedding after the first conv is that the first conv processes the input point cloud to generate a preliminary feature map, where embedding LA can selectively enhance the information in the initial stage of feature extraction, thus strengthening the feature expression of key points and suppressing unimportant point features. The reason for embedding before the final output, after the second conv, is to further optimize feature representation. The LA layer here enhances the interaction of multi-scale features through attention mechanisms, enabling the final output to capture critical information more comprehensively.

Placing the LA module here ensures that the final global features not only contain rich contextual information, but also focus on features that are important for multi-scale local reconstruction, thereby improving the model's performance in downstream tasks. In the final results analysis phase, as we examined, our embedded LA module enabled PointAMaLR to achieve competitive performance in classification and segmentation tasks across multiple datasets.

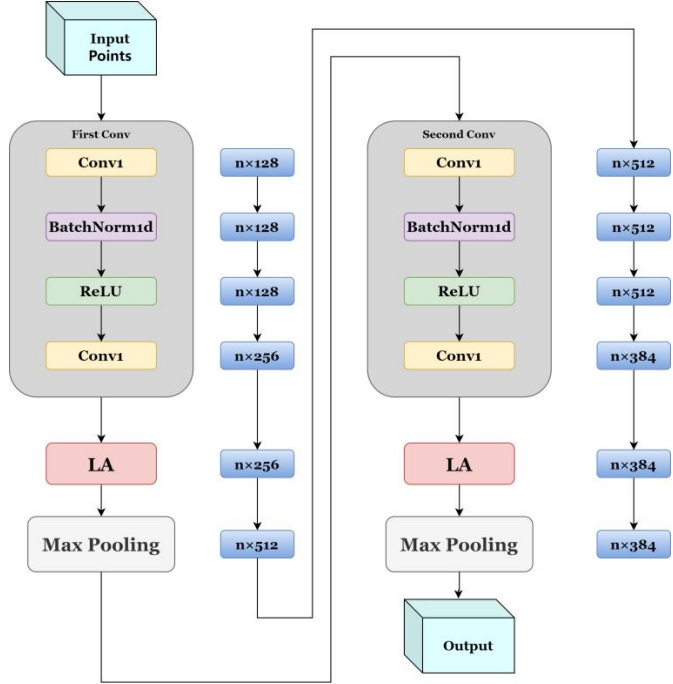

Fig. 2. The structure diagram of the embedding module, where LA refers to the LA module for embedding. After the multi-scale masking, we use a combination of mini-PointNet and the LA module to perform embedding operations on the highest level of the mask, and then send the visible tokens into the subsequent encoder-decoder.

*3.5. LA module*

Accurate representation of three-dimensional shapes is very important for point cloud understanding tasks. The previous work used mini-PointNet to extract the feature markers of the point blocks, which ignored the local geometry of the point cloud. Since point clouds represent three-dimensional shapes, constructing point blocks is a necessary step to study geometric clues in local neighborhoods. In order to strengthen the unique features contained in the focus block and realize the goal of absorbing lower level details in multi-scale local reconstruction, we design the local attention module LA based on the characteristics of point cloud data inspired by the channel attention mechanism (CA), which is specifically applied in the embedding process. The aim is to improve the feature representation ability of point cloud data through local attention mechanism. As shown in Figure 3, the LA module is applied to the input tensor after each convolution operation to enhance the feature representation of the point cloud.

Specifically, the LA module performs average and maximum pooling operations on the input tensors respectively after each convolution operation to obtain two feature representations. The 1D convolution, group normalization, and Sigmoid activation functions are then applied to the two pooled results to process the pooled features, yielding two outputs: $W_x$ and $W_y$. These two outputs are then added together and multiplied by the original input tensor channel by channel to obtain the weighted output tensor. This modular design ensures that the input and output tensor dimensions are aligned, allowing the module to effectively emphasize important features. At the same time, because it emphasizes important features and suppresses unnecessary ones, this allows our model to further improve feature representation. The specific implementation formula for the module is as follows:

$$W_x = \sigma\left(GN\left(Con1\left(AvgPool(Input)\right)\right)\right) \quad (3)$$
$$W_y = \sigma\left(GN\left(Con1\left(MaxPool(Input)\right)\right)\right) \quad (4)$$
$$Output = x \times (W_x + W_y) \quad (5)$$

where input is the tensor obtained from the convolution operation on the upper level of the embedding process. Con1 is a 1D convolution, and GN is a group normalization applied to the convolution output. σ represents the Sigmoid activation function, which further processes the normalized output. The final multiplication stage x represents the original input tensor. This simple module design enables the model to focus on important features in the point cloud by combining multi-scale features of average pooling, which captures global information, and maximum pooling, which focuses on locally significant features. It not only improves feature expression and model flexibility, but also captures multi-scale features effectively, thus avoiding the neglect of lower-level detail features. As we have shown in subsequent experiments, our model effectively improves performance for tasks such as point cloud reconstruction, classification, and segmentation, especially when dealing with complex shapes and multi-scale features to better capture detail and global structure. It is worth noting that our LA module is affected by two hyperparameters λ (window size) and γ (group normalized granularity), mainly by controlling the size of the receptive field and the normalization granularity to optimize the feature capture ability and training stability. The optimal case is set to λ=5 and γ=32.

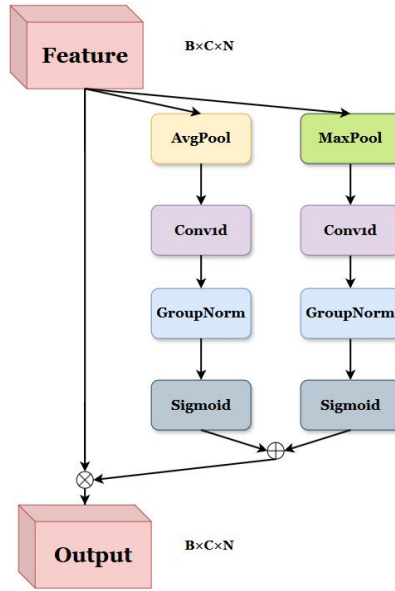

Fig. 3. The structure of LA module. Extract multi-scale features through average pooling and max pooling, respectively. After convolution, group normalization, and activation, sum the results, and finally perform channel-wise weighting with the input tensor to enhance the expressive power of key features.

### 3.6. Encoder-Decoder Design

Based on multi-scale masking, we embed the initial tokens of visible points $P'_1$ into a hierarchical encoder with $S$ stages. Each stage is equipped with $K$ stacked encoder blocks, where each block contains a self-attention layer and a feed-forward network (FFN) composed of multi-layer perceptrons (MLPs). Between every two consecutive stages, we introduce spatial token merging modules to aggregate adjacent visible tokens and enlarge the receptive fields for downsampling the point clouds.

In the decoder stage, thanks to the hierarchical encoder, we obtain the encoded visible tokens $\{T_i^{'}\}_{j=1}^{S}$ for all scales. Starting from the highest $S-$ th scale, we assign a shared learnable mask token to all masked positions $P_S^{''}$ and concatenate them with the visible tokens $T_S^{'}$. We denote them as $\{D_1^{'}, D_1^{''}\}$, with coordinates $\{P_S^{'}, P_S^{''}\}$, serving as the input for the hierarchical decoder. The decoder is designed to be lightweight with $S-1$ th stages, each consisting of only one decoder block, allowing the encoder to embed more semantics of the point clouds. Each decoder block consists of a standard self-attention layer and an FFN. We upsample the point tokens between stages to progressively recover the fine-grained geometries of 3D shapes before reconstruction. The $j-$ th stage of the decoder corresponds to the $(S+1-j)-$ th stage of the encoder, both containing point tokens of the same $(S+1-j)$ scale with feature dimension $C_{S+1-j}$. Between the $(j-1)-th$ and $j-$ th stages (where $1 < j \leq S-1$), we upsample the tokens $\{D_{j-1}^{'}, D_{j-1}^{''}\}$ from the coordinates $\{P_{S+2-j}^{'}, P_{S+2-j}^{''}\}$ to $\{P_{S+1-j}^{'}, P_{S+1-j}^{''}\}$ via the token propagation module. Specifically, we obtain the $k$ nearest neighbors of each point in $\{D_{j-1}^{'}, D_{j-1}^{''}\}$ indexed by $I_{S+2-j}$, and we recover their features through weighted interpolation, resulting in the tokens $\{D_j^{'}, D_j^{''}\}$ for the $j-$ th stage. Our encoder-decoder module is shown in Figure 4.

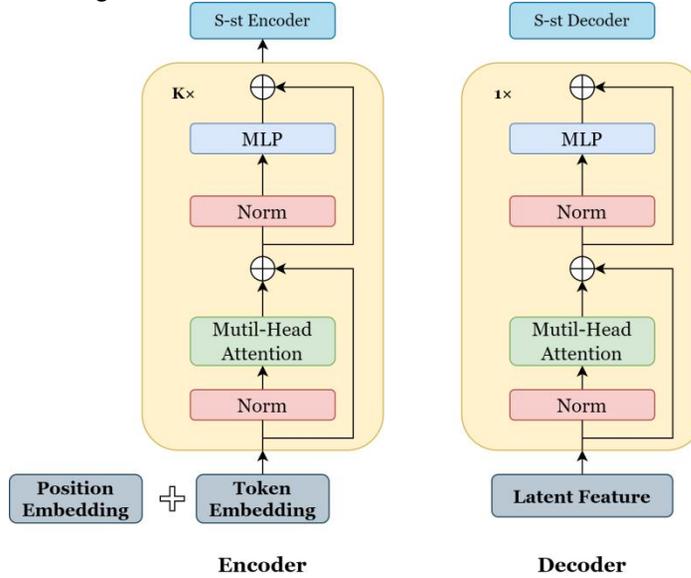

Fig. 4. The structure of S-level encoders and decoders. The encoder processes the input embeddings that have undergone multi-scale masking and embedding operations, which include positional embeddings and token embeddings, to handle potential features, and then the decoder converts these features.

### 3.7. Point Reconstruction

The decoding process reaches its culmination in the reconstruction phase, which is specifically designed to restore the precise coordinates of points within each occluded segment of the point cloud. Following the $S-1$ stage of the decoder, we obtain $\{D_{s-1}^{'}, D_{S-1}^{''}\}$ with coordinates $\{P_2^{'}, P_2^{''}\}$ and reconstruct the masked values using the mask token $D^{"}_{S-1}$. To align with our objective of multi-scale local reconstruction, we not only predict the values at the $0-th$ scale of the input point cloud $P$, but also reconstruct the coordinates of $P^{"}_1$, thereby recovering the masked positions at the first scale $P^{"}_1 \in \mathbb{R}^{N^{"}_1 \times 3}$ from the second scale $P^{"}_2 \in \mathbb{R}^{N^{"}_2 \times 3}$. This is effective because the first scale $\{P'_1, P^{"}_1\}$ can adequately represent the overall 3D shape while retaining sufficient local patterns. For each label in $D^{"}_{S-1} \in R^{N^{"}_2 \times C_2}$, we reconstruct the nearest neighbors indicated by $I_2$ through a reconstruction head composed of a linear projection layer. The loss is then computed using the $l_2$ Chamfer Distance. Drawing on insights from prior research, we utilize LCD for supervision, enabling a pure masked autoencoding approach for self-supervised pre-training. The specific formula is as follows:

$$\widetilde{\widetilde{P}}^{"}_{2\to1} = Linear\left(D^{"}_{S-1}\right) \quad (6)$$

where $\widetilde{\widetilde{P}}^{"}_{2\to1} \in \mathbb{R}^{N^{"}_2 \times k \times 3}$. The loss is defined as:

$$\mathcal{L} = ChamferDistance\left(P^{"}_{2\to1}, \widetilde{\widetilde{P}}^{"}_{2\to1}\right) \quad (7)$$

where $\widetilde{\widetilde{P}}^{"}_{2\to1}$ and $P^{"}_{2\to1}$ represent the predicted and ground-truth reconstruction coordinates from the second scale to the first scale. We employ only the LCD for supervision to facilitate a purely masked autoencoding process for self-supervised pre-training. The specific formula of $l_2$ ChamferDistance is as follows:

$$\mathcal{L}\left(P^{"}_{2\to1}, \widetilde{\widetilde{P}}^{"}_{2\to1}\right) \& = \frac{1}{|P^{"}_{2\to1}|}\sum_{p\in P^{"}_{2\to1}} \min_{\widetilde{p}\in\widetilde{\widetilde{P}}^{"}_{2\to1}} \|p-\widetilde{p}\|^2_2 + \frac{1}{|\widetilde{\widetilde{P}}^{"}_{2\to1}|}\sum_{\widetilde{p}\in\widetilde{\widetilde{P}}^{"}_{2\to1}} \min_{p\in P^{"}_{2\to1}} \|\widetilde{p}-p\|^2_2$$

$$(8)$$

## 4. Experiments

To evaluate the effectiveness of our approach, we performed a comprehensive series of experiments. This section outlines the datasets and experimental settings, the pre-training strategies employed, and the performance metrics achieved in downstream tasks. Additionally, we declare that all experiments were conducted using a Titan RTX graphics card.

*4.1. Datasets*

Our experiments used four popular point cloud standard datasets: ModelNet40[33], ScanObjectNN[34], ShapeNet and S3DIS[35].

ModelNet40. The ModelNet40 dataset is a synthetic dataset comprising 40 categories with a total of 12,311 models. It is divided into a training set with 9,843 samples and a test set with 2,468 samples. Derived from CAD models, ModelNet40 is characterized by its noise-free data. Each sample consists of 1,024 points, and data augmentation techniques are applied to these points as inputs to the network.

ScanObjectNN. The ScanObjectNN dataset is a real-world dataset containing 2,902 objects scanned from various environments. The training dataset encompasses 2,321 point clouds, while the test dataset contains 581 point clouds. This dataset presents significant challenges due to the presence of missing parts, noise points, and background interference, making classification a complex task. For the classification task, each point cloud is processed by randomly sampling 1,024 points and normalizing the coordinates of each point.

ShapeNet. The ShapeNet dataset is known for its extensive variety of CAD models, covering 16 categories such as cups, knives, hats, and more. Each sample within this dataset comprises 2-6 distinct parts, totaling 55 different parts. The training set includes 14,007 samples, while the test set contains 2,874 samples.

S3DIS. The S3DIS dataset is a 3D point cloud dataset that includes six large indoor areas with a total of 271 rooms. Each point in the scene point cloud is labeled with one of 13 semantic categories, including ceiling, floor, wall, beam, column, window, door, table, chair, sofa, bookshelf, whiteboard, and clutter, etc. The point cloud data of each room is divided into 1-meter by 1-meter blocks, with each block containing 4096 points. Each point is represented by 9-dimensional data, including XYZ coordinates, RGB color information, and normalized coordinates relative to the room position (values range from 0 to 1). This dataset is widely used for tasks such as 3D semantic segmentation and point cloud classification and is an important resource in the field of indoor scene understanding.

*4.2. Pretrain Setting*

We pre-trained our PointAMaLR model on the ShapeNet dataset, which includes 57,448 synthetic 3D shapes across 55 categories. For this pre-training, we extracted 2,048 points from each 3D shape, treating these points as the fundamental data units for our model. The network was trained for 300 epochs with a batch size of 64, using AdamW as the optimizer. We employed the CosLR scheduler, which adjusts the learning rate according to the cosine function, potentially enhancing convergence properties. The initial learning rate and weight decay were set to 0.0001 and 0.05, respectively. The first 10 training epochs were allocated for preheating, a strategy that gradually increases the learning rate from a minimal value to the initial rate, promoting faster and more stable convergence. Data augmentation techniques, including random scaling and translation, were applied during pre-training to enhance the dataset. In the pre-training period, we configured our model with 3 stages (S=3), comprising a 3-stage encoder and a two-level decoder for hierarchical learning. Each encoder stage consists of 5 blocks. We assigned different k-values for different scales of k-NN searches: {16, 8, 8}. For a 3-scale point cloud, the point and token dimensions were set to {512, 256, 64} and {96, 192, 384}, respectively. We employed masking on the highest proportion of point clouds and equipped each decoder layer with 6 attention heads, which is crucial for the accurate reconstruction of the masked points.

*4.3. Results on Downstream Tasks*

*4.3.1. Object classification on ModelNet40*

For shape classification tasks, we implemented fine-tuning strategies on the ModelNet40 datasets. In these tasks, the input point cloud for each model was uniformly set to 1,024 points. The training process spanned over 300 epochs, with each batch comprising 32 samples. We selected AdamW as the optimizer, setting the learning rate to 0.0005 and the weight decay (attenuation factor) to 0.05. We employed the Cosine Annealing Learning Rate scheduler, CosLR, to manage the training process. The Transformer encoder architecture was kept consistent with that used in the pre-training phase. To ensure the integrity of the experiment, a uniform number of 1,024 input points was maintained across all configurations. Our experimental results are presented in Table 1. Notably, to ensure a fair comparison, we reproduced groundbreaking works such as Point-BERT and Point-MAE, which are foundational to the masked autoencoder concept, in our experimental environment using their official code and pre-trained models. The term 'Rep' denotes the replicated results of these models in our setup.

As depicted in Table 1, our method is benchmarked against a range of classical supervised learning and self-supervised learning methods for point cloud shape classification.

Analysis of the data presented in Table 1 reveals that PointAMaLR surpasses most supervised learning techniques. Specifically, our approach achieves accuracy improvements of 4.24%, 2.74%, and 0.54% over PointNet, PointNet++, and DGCNN, respectively. When compared to Transformer-based supervised learning methods, PointAMaLR exhibits a 2.04% higher accuracy than Transformer and a 0.24% advantage over PCT. Furthermore, in the realm of self-supervised learning, our approach demonstrates superiority over most existing methods. Its accuracy is 0.34% higher than STRL, and 0.44% greater when compared to OcCo. In direct comparison with Point-BERT and Point-MAE, PointAMaLR outperformed Point-BERT, Point-BERT (R) and Point-MAE (R) by 0.24%, 0.34% and 0.33%, respectively.

Table 1: Object Classification on ModelNet40[T] represents the model is based on modified Transformers. [ST] represents the standard Transformers models

| Training category | Methods | Accuracy |
| --- | --- | --- |
| Supervised methods | PointNet[1] | 89.2% |
|  | PointNet++[23] | 90.7% |
|  | PointCNN[36] | 92.5% |
|  | DGCNN[24] | 92.9% |
|  | RS-CNN[37] | 92.9% |
|  | [T]PCT[38] | 93.2% |
|  | [T]Point Transfomer[26] | 93.7% |
| Self-supervised methods | OcCo[31] | 93.0% |
|  | STRL[30] | 93.1% |
|  | [ST]Point-BERT[16] | 93.2% |
|  | [ST]Point-BERT (Rep) | 93.1% |
|  | [ST]Point-MAE[17] | 93.8% |
|  | [ST]Point-MAE (Rep) | 93.11% |
|  | [ST]PointGAME[19] | 93.35% |
|  | Ours | **93.44%** |

PointAMaLR demonstrates significant performance improvements, consistently outperforming various mainstream supervised and self-supervised methods. This advantage stems from its enhanced feature representation capacity, multi-scale feature capture, self-supervised learning design, and module adaptability. By incorporating a multi-scale local attention mechanism, PointAMaLR more accurately captures local features, improving its performance on complex data. Moreover, PointAMaLR achieves higher accuracy through multi-scale feature-based self-supervised learning while maintaining efficiency. Notably, compared to Point-MAE, PointAMaLR outperforms our locally reproduced results of Point-MAE, further validating the advantage of multi-scale local reconstruction strategies in point cloud data representation. Overall, these designs enable PointAMaLR to achieve competitive performance gains in point cloud classification tasks, offering stronger generalization and adaptability than existing methods.

The experimental results underscore the high efficiency of our pre-trained model and its competitive performance in the domain of shape classification. At the same time, we used t-SNE to visualize the feature distribution in the experiment, and the results are shown in Figure 5.

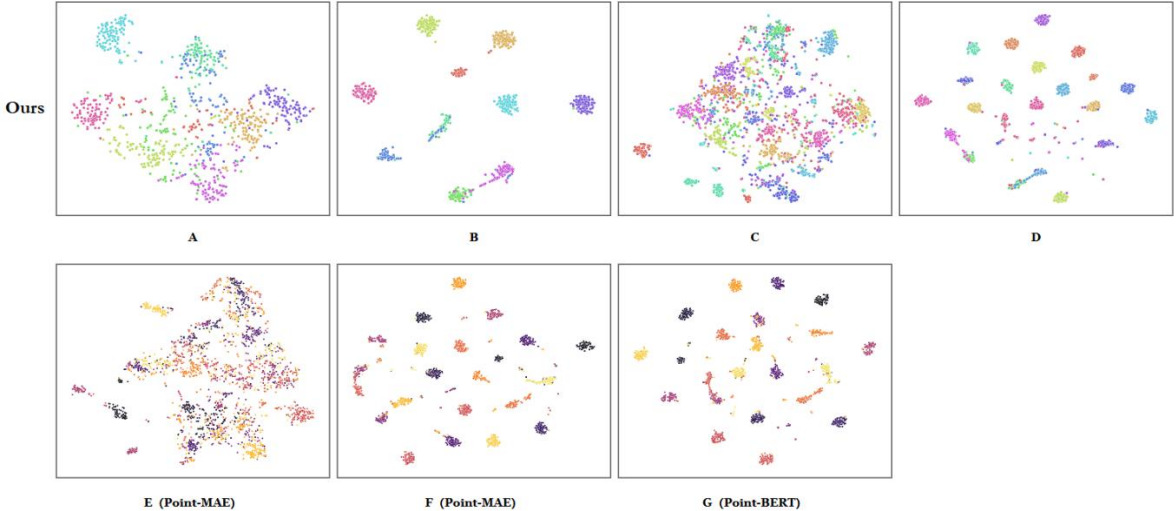

Fig. 5. Visualization of feature distribution. We present t-SNE visualizations of our model, Point-MAE, and feature vectors for Point-BERT learning. (A) Our model is pre-trained with ModelNet10. (B) Our model has been fine-tuned with ModelNet10. (C) Our model is pre-trained with ModelNet40. (D) Our model is fine-tuned with ModelNet10. (E) Point-MAE after ModelNet40 pre-training. (F) Point-MAE after ModelNet40 fine-tuning. (G) Point-BERT after ModelNet40 fine-tuning.

## 4.3.2. Object classification on Real-World Dataset

Table 2: Object Classification on ScanObjectNN

| Methods | OBJ-BG | OBJ-ONLY | RB-T50-RS |
|---|---|---|---|
| PointNet[1] | 73.3 | 79.2 | 68 |
| SpiderCNN[39] | 77.1 | 79.5 | 73.7 |
| PointNet++[23] | 82.3 | 84.3 | 77.9 |
| DGCNN[24] | 82.8 | 86.2 | 78.1 |
| BGA-DGCNN[34] | - | - | 79.7 |
| PRANet[40] | - | - | 81.0 |
| Transformer[16] | 79.86 | 80.55 | 77.24 |
| Transformer + OcCo[31] | 84.85 | 85.54 | 78.79 |
| Point-BERT[16] | 87.43 | 88.12 | 83.07 |
| Point-BERT (Rep) | 87.43 | 86.91 | 83.10 |
| Point-MAE[17] | 90.02 | 88.29 | 85.18 |
| Point-MAE (Rep) | 88.98 | 88.29 | 84.31 |
| Point-M2AE[32] | 91.22 | 88.81 | **86.43** |
| Point-M2AE(Rep) | 90.31 | 88.47 | 84.97 |
| Ours | **91.74** | **89.34** | 85.34 |

Our pre-trained model, which is solely based on the object models from the ShapeNet dataset without incorporating scenario features, undergoes rigorous testing for robustness and generalization on the challenging real-world dataset, ScanObjectNN. To effectively handle noisy spatial structures present in ScanObjectNN, we have adjusted the k-values for k-NN to {32, 16, 16}, thereby enhancing the encoding of local patterns with broader acceptance domains.

In line with protocols established by previous studies, we conducted experiments on three primary variants: OBJ-BG, OBJ-ONLY, and PB-T50-RS. The results of these assessments are detailed in Table 2.

From the data in Table 2, it can be seen that our PointAMaLR not only achieves high performance in the three variants of ScanObjectNN, but also in the performance of the OBJ-BG and OBJ-ONLY variants compared with that of the masked autoencoder architecture PointBETR, Point-MAE, Point-MAE. This shows that our local attention-directed multi-scale coding method has significant advantages in complex scenes. In addition, given that ScanObjectNN is a real-world dataset with a large semantic gap from the pre-trained ShapeNet dataset, the powerful performance of PointAMaLR proves that it is powerful and can be transferred from different domains. This highlights the exceptional representation learning ability of our pre-trained model and its remarkable generalization across various domains.

*4.3.3. Part Segmentation*

Table 3: Part Segmentation on ShapeNetPart

| Methods | $mIoU_c$ | $mIoU_i$ | Airplane / lamp | bag / laptop | cap / motor | car / mug | chair / pistol | earphone / rocket | guitar / skateboard | knife / table |
|---|---|---|---|---|---|---|---|---|---|---|
| PointNet[1] | 80.39 | 83.7 | 83.4 / 80.8 | 78.7 / 95.3 | 82.5 / 65.2 | 74.9 / 93 | 89.6 / 81.2 | 73.0 / 57.9 | 91.5 / 72.8 | 85.9 / 80.6 |
| PointNet++[23] | 81.85 | 85.1 | 82.4 / 83.7 | 79.0 / 95.3 | 86.7 / 71.6 | 77.3 / 94.1 | 90.8 / 81.3 | 71.8 / 58.7 | 91.0 / 76.4 | 85.9 / 82.6 |
| DGCNN[24] | 82.33 | 85.2 | 84.0 / 82.8 | 83.4 / 95.7 | 86.7 / 66.3 | 77.8 / 94.9 | 90.6 / 81.1 | 74.7 / 63.5 | 91.2 / 74.5 | 87.5 / 82.6 |
| Transformer[16] | 83.42 | 85.1 | 82.9 / 85.3 | 85.4 / 95.6 | 87.7 / 73.9 | 78.8 / 94.9 | 90.5 / 83.5 | 80.8 / 61.2 | 91.1 / 74.9 | 87.7 / 80.6 |
| Point-BERT[16] | 84.11 | 85.6 | 84.3 / 85.2 | 84.8 / 95.6 | 88.0 / 75.6 | 79.8 / 94.7 | 91.0 / 84.3 | **81.7** / 63.4 | 91.6 / 76.3 | 87.9 / 81.5 |
| Point-MAE[17] | 84.19 | 86.1 | 84.3 / **86.1** | 85.0 / 96.1 | **88.3** / 75.2 | 80.5 / 94.6 | 91.3 / 84.7 | 78.5 / 63.5 | 92.1 / **77.1** | 87.7 / 82.4 |
| **Ours** | **85.00** | **86.49** | **85.3** / 86.0 | **87.2** / **96.2** | 88.0 / **77.0** | **81.3** / **95.3** | **91.7** / **84.8** | 77.2 / **65.6** | **92.3** / 76.2 | **88.7** / 82.6 |

Partial segmentation of point cloud objects poses a significant challenge in assessing the proficiency of pre-trained models, which are tasked with assigning category labels to individual points. We conducted an evaluation of PointAMaLR on the ShapeNetPart dataset for partial segmentation, a task that demands the model to predict partial labels for each point and necessitates a nuanced understanding of local patterns. To ascertain the effectiveness of our pre-training in capturing both high-level semantics and fine-grained details, we employed exceedingly simple segmentation heads. Drawing on previous research, our method samples 2,048 points per object for input. Subsequently, we concatenate the upsampled 3-scale features for each point and utilize stacked linear projection layers to predict partial labels. Table 3 shows our experimental results in detail.

PointAMaLR's mIoU scores are 85.00% and 86.49%, respectively, both of which exceed those of Point-BERT and Point-MAE. Notably, our method demonstrates outstanding performance across twelve distinct categories: Airplane, bag, car, chair, guitar, knife, laptop, motor, mug, pistol, rocket and table. These results highlight the significant advantages of PointAMaLR in feature extraction and multi-scale reconstruction. The model's success can be attributed to its attention-guided multi-scale local reconstruction approach, which effectively captures the subtle features of different categories, enhancing performance across multiple categories. By better representing local information and contextual relationships, PointAMaLR excels in handling categories with complex geometric structures. However, the performance in certain categories did not reach leading levels, which may involve several factors. Firstly, the feature complexity of different categories may vary; some categories may exhibit higher variability or more intricate structures, making it challenging for the model to effectively extract their features.

*4.3.4. Few-shot Learning*

Table 4: Few-shot object classification on ModelNet40. We conduct 10 independent experiments for each setting and report mean accuracy (%) with standard deviation.

| Methods | 5-way,10-shot | 5-way,20-shot |
| --- | --- | --- |
|  | 10-way,10-shot | 10-way,20-shot |
| DGCNN-rand[31] | 31.6 ± 2.8 | 40.8 ± 4.6 |
|  | 19.9 ± 2.1 | 16.9 ± 1.5 |
| DGCNN-OcCo[31] | 90.6 ± 2.8 | 92.5 ± 1.9 |
|  | 82.9 ± 1.3 | 86.5 ± 2.2 |
| Transformer-rand[16] | 87.8 ± 5.2 | 93.3 ± 4.3 |
|  | 84.6 ± 5.5 | 89.4 ± 6.3 |
| Transformer-OcCo[16] | 94.0 ± 3.6 | 95.9 ± 2.3 |
|  | 89.4 ± 5.1 | 92.4 ± 4.6 |
| Point-BERT[16] | 94.6 ± 3.1 | 96.3 ± 2.7 |
|  | 91.0 ± 5.4 | 92.7 ± 5.1 |
| Point-MAE[17] | 96.3 ± 2.5 | 97.8 ± 1.8 |
|  | 92.6 ± 4.1 | 95.0 ± 3.0 |
| PointGAME[19] | 96.1 ± 3.1 | 97.9 ± 1.4 |
|  | 90.3 ± 5.1 | 93.8 ± 2.5 |
| Point-M2AE[32] | 96.8 ± 1.8 | 98.3 ± 1.4 |
|  | 92.3 ± 4.5 | 95.0 ± 3.0 |
| Ours | 96.4 ± 2.1 | 98.4 ± 1.3 |
|  | 92.6 ± 4.3 | 94.8 ± 2.2 |

Following the previous work[22, 23], we used the pre-trained model on ModelNet40 to conduct a small sample learning experiment. We adopt an n-way, M-shot setup, where n represents the number of categories randomly selected from the dataset and m represents the number of objects randomly sampled for each category. We trained with n×m objects, and for evaluation purposes, we randomly selected 20 objects from each category that had not been seen during the training.

In the experiment, we set n∈5,10 m∈10,20, and show the results in Table 4. Experimental data show that our model still achieves competitive results in few-shot learning.

### 4.3.5. 3-D scene segmentation

Scene segmentation is a challenging task, especially in large-scale 3D scenes, as it demonstrates a model's ability to understand contextual semantics and complex local geometric relationships. We performed 3D scene segmentation on the S3DIS dataset, which provides dense semantic tags for the point cloud. The results can be observed in Table 5. PointAMaLR achieved competitive results in scene segmentation scenarios, outperforming Point-MAE by 0.2% and 1.2% in mean accuracy (mAcc) and intersection over union (mIoU), respectively, and is only 0.9% lower than PCP-MAE in mean accuracy (mAcc). In Figure 6, we show the visualized results of the 3D scene segmentation task. As you can see, our network is able to output smooth and accurate predictions and is robust to missing points and occlusion.

Table 5: Segmentation results on S3DIS Area 5: Mean accuracy (mAcc(%)) and Intersection over Union (mIoU(%)) for semantic segmentation.

| Methods | Semantic Segmentation | |
|---|---|---|
| | mAcc | mIoU |
| PointNet[1] | 49.0 | 41.1 |
| PointNet++[23] | 67.1 | 53.5 |
| Transformer[16] | 68.6 | 60.0 |
| Point-MAE[17] | 69.9 | 60.8 |
| PCPMAE[20] | **71.0** | 61.3 |
| Ours | 70.1 | **62.0** |

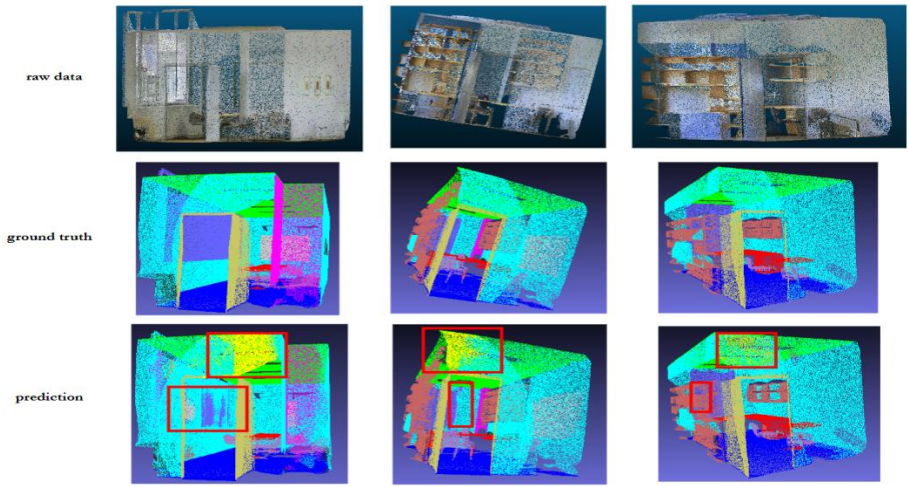

Fig. 6. Visualization results for the 3D scene segmentation task. The top row shows the input raw point clouds, the middle row displays the original point clouds with color from the same viewpoint, and the bottom row shows the output predicted semantic segmentation results from the same viewpoint. The red boxes mark areas where our predictions have minor flaws for a fair presentation.

*4.4. Ablation Study*

In this section, we will explore and perform a series of ablation experiments to systematically validate the critical role of our proposed LA module in multi-scale point cloud local reconstruction tasks. In addition, we will pay special attention to the effect of mask ratio on the learning ability of the model.

*4.4.1. The effect of LA attention module*

To verify the role of the LA module, we designed a detailed ablation experiment in the shape classification task of ModelNet40 to evaluate its effect on the model performance by removing different branches of the module and removing the module completely. In ShapeNetPart's part segmentation task, only two cases of fully removed modules and fully retained modules were compared to evaluate the overall impact. The results in Table 6 and Table 7 show that the removal of different branches leads to a decrease in classification performance. Specifically, when the maximum pooled branch was removed, the accuracy decreased by 0.73%; When the average pooling branch is removed, the accuracy decreases by 0.88%. When the two pooling branches were removed, the accuracy dropped by 1.7%, while the indicators in the part-splitting experiment dropped by 0.7 and 0.4. These results show that the LA module we designed plays a crucial role in capturing multi-scale features. By combining the two branches, LA module can effectively extract global and local features and enhance the classification ability of the model. After these branches are removed, the model is limited in its ability to integrate information at different scales, resulting in a significant decline in performance. These results highlight the important role of LA modules in improving the overall performance and accuracy of models.

Table 6: The impact of the LA module on shape classification on ModelNet40

| $W_x$ | $W_y$ | ACC |
|---|---|---|
| ✓ | ✓ | **93.44%** |
| ✓ | ✗ | 92.71% |
| ✗ | ✓ | 92.56% |
| ✗ | ✗ | 92.13% |

Table 7: The impact of the LA module on part segmentation in ShapeNetPart

| Methods | | $mIoU_c$ | $mIoU_i$ | Airplane | bag | cap | car | chair | earphone | guitar | knife |
|---|---|---|---|---|---|---|---|---|---|---|---|
| $W_x$ | $W_y$ | | | lamp | laptop | motor | mug | pistol | rocket | skateboard | table |
| ✗ | ✗ | 84.31 | 86.1 | 84.4 | 86.6 | 88.0 | 80.3 | 91.4 | 73.7 | 91.8 | 88.2 |
| | | | | 86.2 | 95.7 | 74.0 | 95.0 | 84.2 | 64.1 | 76.9 | 82.2 |
| ✓ | ✓ | 85.0 | 86.49 | 85.3 | 87.2 | 88.0 | 81.3 | 91.7 | 77.2 | 92.3 | 88.7 |
| | | | | 86.0 | 96.2 | 77.0 | 95.3 | 84.8 | 65.6 | 76.2 | 82.6 |

*4.4.2. The effect of Masking Strategy*

Table 8 illustrates the performance of PointAMaLR under various mask settings, indicating that a mask ratio of 60% yields the optimal results. Here, we use the parameter $\mu$ to represent the mask ratio. This finding suggests that at this mask ratio, the model faces an appropriate level of challenge, which facilitates the learning of meaningful representations during self-supervised pre-training. The appropriate mask ratio is closely related to multi-scale local reconstruction, as it effectively enhances the model's ability to extract information at different scales. A mask ratio that is too low may not provide sufficient challenge, preventing the model from effectively extracting robust features, while a ratio that is too high could hinder learning by causing the model to lose important information necessary for reconstruction. The 60% mask ratio strikes a balance, allowing the model to focus on reconstructing the unmasked features while still benefiting from the context provided by the masked points. This setup not only improves the model's ability to reconstruct local features but also enhances its performance in multi-scale feature fusion. Therefore, this optimal setting supports the model's ability to generalize well across different tasks and underscores the importance of carefully adjusting the masking strategy in self-supervised learning scenarios.

Table 8: Classification of objects with different mask rate distributions on ModelNet40

| $\mu$ | *Accuracy* |
|---|---|
| 0.9 | 92.91% |
| 0.8 | 93.27% |
| 0.7 | 92.79% |
| **0.6** | **93.44%** |
| 0.5 | 92.88% |

Since our PointAMaLR focuses on the local reconstruction of multi-scale point clouds guided by attention, we place special emphasis on the final reconstruction quality. Therefore, we visualized the reconstruction results for different categories after applying the optimal mask ratio, as shown in Figure 7.

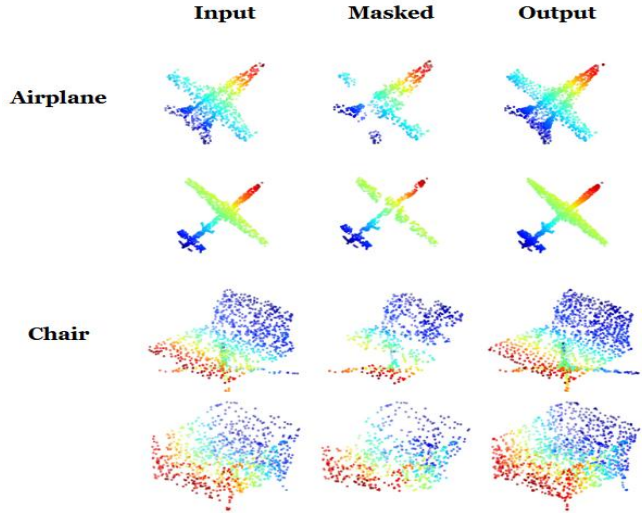

Fig. 7. Visualization of point cloud reconstruction.

*4.4.3. The effect of hyperparameters $\lambda$ and $\gamma$*

To validate the impact of the LA module's hyperparameters on model performance, we conducted an ablation experiment based on $\lambda$ and $\gamma$. Specifically, we tested different settings of $\lambda$ and $\gamma$ to assess their contributions to model performance.

Table 9: The Impact of hyperparameters $\lambda$ and $\gamma$ on shape classification on ModelNet40

|         | $\lambda$ | $\gamma$ | ACC |
|---------|---|----|---------|
| Model-A | **5** | **32** | **93.44%** |
| Model-B | 5 | 16 | 92.67% |
| Model-C | 7 | 32 | 93.13% |
| Model-D | 7 | 6  | 92.34% |

In the experiment, we first set $\lambda = 5$ and $\gamma = 32$ as the optimal configuration, which serves as the baseline result. In addition, we set up a total of four models, of which Model-A is the baseline model. Then, the configuration of Model-B is $\lambda = 5$ and $\gamma = 16$; The configuration of Model-C is $\lambda = 7$ and $\gamma = 32$; The configuration of Model-D is $\lambda = 7$ and $\gamma = 16$. In Table 9, we show the accuracy results of classifying Model40 datasets after different configuration models are trained under the same training environment.

The experimental results show that the optimal configuration ($\lambda = 5$, $\gamma = 32$) provides the best balance in capturing local features and multi-scale information, thereby improving overall model performance. When the window size or group normalization granularity deviates from the optimal value, the model's performance declines to varying degrees, demonstrating the critical role of hyperparameter selection in the LA module's performance.

## 5. Conclusion and limitations

In this paper, we present PointAMaLR, an innovative self-supervised learning framework for point cloud processing that advances feature representation and processing accuracy through attention-guided multi-scale local reconstruction. PointAMaLR implements hierarchical reconstruction across multiple local regions, with lower layers dedicated to fine-scale feature restoration and upper layers focusing on coarse-scale reconstruction, thereby enabling sophisticated inter-patch interactions. To augment semantic feature understanding, we incorporate a Local Attention (LA) module within each layer. Comprehensive experimental evaluations demonstrate the model's effectiveness during pre-training and its robust efficiency and generalization capabilities in downstream tasks, including point cloud classification and segmentation. PointAMaLR achieves exceptional classification accuracy and superior reconstruction quality on benchmark datasets such as ModelNet and ShapeNet, while demonstrating competitive performance on real-world datasets like ScanObjectNN, even attaining state-of-the-art results on two of its variants. Furthermore, the framework exhibits strong performance and generalization abilities in large-scale scene segmentation tasks, as evidenced by its results on the S3DIS dataset.

**Limitations.** Nevertheless, we acknowledge the inherent challenges posed by the irregular nature of point cloud data. The current framework shows a performance gap in high-precision classification when compared to the most advanced technologies, which we attribute primarily to variations in the reconstruction process of masked point clouds. Addressing these limitations will constitute a primary focus of our future research. Specifically, we plan to enhance high-level cloud representation learning capabilities by refining our approach with dynamic masking strategies, which we anticipate will significantly improve both point cloud reconstruction quality and downstream task performance.

**Declaration of competing interest**

The authors declare that they have no known competing financial interests or personal relationships that could have appeared to influence the work reported in this paper.

**Data availability**

All data are from publicly available datasets and are freely available on the Internet.